\documentclass[prc, floatfix, twoside, nofootinbib,showpacs,twocolumn]{revtex4-1}
\usepackage[english]{babel}
\usepackage[bookmarks,pdfhighlight=/O,colorlinks=false,pdfstartview=FitH]{hyperref}
\usepackage{amsmath,amssymb,mathrsfs,slashed,bm}
\usepackage[dvipsnames]{xcolor}
\usepackage{graphicx}
\usepackage{epstopdf}
\usepackage{natbib}

 \newcommand{\GeV}{\mathrm{GeV}}

\newcommand{\fm}{\mathrm{fm}}

\def\snn{\sqrt {s_{NN}}}

\def\x{{\boldsymbol x}}

\begin{document}

\preprint{}

\title{Temperatures and chemical potentials at kinetic freeze-out in relativistic heavy ion collisions from coarse grained transport simulations\\}

\author{Gabriele Inghirami$^{1,2}$}
\author{Paula Hillmann$^{3,4,5,6}$}
\author{Boris Tom\'a\v{s}ik$^{7,8}$}
\author{Marcus Bleicher$^{3,4,5,6}$}

\affiliation{$^{1}\,$University of Jyv\"askyl\"a,  Department of Physics,
	P.O. Box 35, FI-40014 University of Jyv\"askyl\"a, Finland\\
$^{2}\,$Helsinki Institute of Physics, P.O. Box 64, FI-00014 University of Helsinki, Finland\\
$^{3}\,$Frankfurt Institute for Advanced Studies (FIAS),Ruth-Moufang-Str. 1, 60438 Frankfurt am Main, Germany\\
$^{4}\,$Institut f\"ur Theoretische Physik, Johann Wolfgang Goethe-Universit\"at, Max-von-Laue-Str. 1, 60438 Frankfurt am Main,
  Germany\\
$^{5}\,$GSI Helmholtzzentrum f\"ur Schwerionenforschung GmbH, Planckstra{\ss}e 1, 64291 Darmstadt, Germany\\
$^{6}\,$John von Neumann Institute for Computing, Forschungszentrum J\"ulich, 52425 J\"ulich, Germany\\
$^{7}\,$Fakulta pr\'irodn\'ych vied, Univerzita Mateja Bela, 97401 Bansk\'a Bystrica, Slovakia\\
$^{8}\,$Fakulta jadern\'a a fyzik\'aln\v{e} in\v{z}en\'yrsk\'a, \v{C}esk\'e vysok\'e u\v{c}en\'i technick\'e v Praze, 11519 Prague, Czechia}

\date{\today}

\begin{abstract}
	
Using the UrQMD/coarse graining approach we explore the kinetic freeze-out stage in central Au + Au collisions at various energies. These studies allow us to obtain detailed information on the thermodynamic properties (e.g. temperature and chemical potential) of the system during the kinetic decoupling stage. We explore five relevant collision energies in detail, ranging from $\sqrt{s_{NN}}=2.4\,\GeV$ (GSI-SIS) to $\sqrt{s_{NN}}=200\,\GeV$ (RHIC). By adopting a standard Hadron Resonance Gas equation of state, we determine the average temperature $\langle T \rangle$ and the average baryon chemical potential $\langle\mu_{\mathrm{B}}\rangle$ on the space-time hyper-surface of last interaction. The results highlight the nature of the kinetic freeze-out as a continuous process. This differential decoupling is an important aspect often missed when summarizing data as single points in the phase diagram as e.g. done in Blast-Wave fits. We compare the key properties of the system derived by using our approach with other models and we briefly review similarities and differences. 

\end{abstract}


\maketitle
\section*{}
This is the Accepted Manuscript version of an article accepted for publication in \emph{Journal of Physics G: Nuclear and Particle Physics}.  IOP Publishing Ltd is not responsible for any errors or omissions in this version of the manuscript or any version derived from it.  The Version of Record is available online at\\ \url{https://doi.org/10.1088/1361-6471/ab53f4}.
\section{Introduction}
Heavy ion collisions at ultra-relativistic energies have provided strong evidences~\cite{Adams:2005dq,Heinz:2002gs,Blaizot:1996nq,Bass:1998vz, Eskola:1999sx} for a novel phase of Quantum-Chromo-Dynamic (QCD) matter. This novel state of deconfined matter~\cite{Hagedorn:1965st,Matsui:1986dk} is called the (strongly interacting) Quark-Gluon Plasma (QGP). A large variety of approaches~\cite{Philipsen:2012nu,Rapp:2019bxp,Song:2015sfa,Bleicher:2014wua,Pang:2019int,Plumberg:2016sig,Petersen:2018jag,Gallmeister:2017ths,Karpenko:2018erl,Nahrgang:2018afz,Schenke:2019ruo,Wang:2019vaz,Tu:2019mso} have been developed to study the properties of this QCD-medium, allowing to test in detail our understanding of the laws of nature at the subatomic scale. Unfortunately, the tiny dimensions of the QGP system under investigation and its extremely fast evolution make it inaccessible to direct measurements. Therefore, one is constrained - even with the most advanced experimental apparatus - to the detection of hadrons and their momentum distributions at distances many orders of magnitudes larger than the typical size of the colliding ions. Dynamical modelling, however, opens a key hole to explore the intriguing and exciting phenomena happening in the early stages of the collision. Nevertheless, it is clear that this indirect view relies on the quality of the model to consistently and accurately reconstruct the relevant dynamics and phases of the collision from hadron formation to their detection.

In this work we want to explore the systems properties during the decoupling stage of the evolution. A similar analysis was e.g. done in \cite{Knoll:2008sc} in a more ab-initio fashion, however with less realistic initial conditions and only with a schematic expansion and more phenomenologically in \cite{Bellwied:2000mi}. Thus, we focus on the last stage of a heavy ion collision event, the so called \emph{kinetic} or \emph{thermal freeze-out}~\cite{Magas:1999yb,Molnar:2005gx}, when the hadrons stop to interact with each other and their momentum distribution does not change anymore. This condition is different from the so-called \emph{chemical freeze-out}~\cite{Cleymans:1999st,Florkowski:1999pz,BraunMunzinger:2003zz}, which, instead, refers to the ceasing of the inelastic scatterings and the stabilization of the abundances of the hadronic species. Although single freeze-out models have been proposed~\cite{Broniowski:2001we,Broniowski:2001uk,Baran:2003nm} and some models estimate~\cite{Prorok:2007xp} a chemical freeze-out temperature $T_{ch}$ close to the kinetic freeze-out temperature $T_{kin}$, the two phenomena are conceptually different~\cite{Chatterjee:2015fua,Becattini:2017pxe}. $T_{ch}$ is tightly connected with the QGP phase transition~\cite{Heinz:2006ur,Bass:1999tu}, it depends on the collision energy~\cite{Cleymans:2005xv}, but not on the collision centrality class~\cite{Cleymans:2004pp} and it has common features in different systems well explained by statistical thermal hadronization models~\cite{Andronic:2017pug,Becattini:1997rv,BraunMunzinger:2000px,Becattini:2008tx,Andronic:2008ev}. On the other hand, $T_{kin}$ is more related to the dynamics of the system~\cite{Magas:1999yb}. The results delivered by the recent versions of the multi-source thermal model~\cite{Wei:2016ihj} and, even more, by the Blast-Wave model~\cite{Schnedermann:1993ws} heavily depend on the assumptions about the kinematic properties of the system. In particular, in the Blast-Wave model the kinetic freeze-out temperature, the baryon chemical potential and the transverse velocity are parameters obtained by a fit to a certain phase-space density distribution of hadrons~\cite{Melo:2015wpa,Abelev:2008ab,Kumar:2014tca,Adamczyk:2017iwn,Melo:2019temp}. The Blast-Wave model can be quite sophisticated~\cite{Rode:2018hlj} and may take into account the anisotropy of the system~\cite{Cimerman:2017lmm}, but, in any case, it is an approach based on the direct evaluation of \emph{macroscopic} quantities fitted to experimental data.

In this paper we adopt a different perspective. We exploit a \emph{microscopic} description of the system given by the numerical transport code UrQMD~\cite{Bass:1998ca,Bleicher:1999xi} and then we associate to the kinetic freeze-out condition the corresponding macroscopic quantities by using a coarse-graining approach~\cite{Huovinen:2002im,Endres:2014zua}. We define the kinetic freeze-out microscopically as the time and the position in space of the last interaction of a hadron, including not only scatterings, but also decays by strong interaction. Therefore, within the present framework, the freeze-out coordinates are given by the dynamics and cross sections of the UrQMD simulation. To relate these freeze-out coordinates to the thermal properties at this space-time point, we compute in a second step the average net-baryon current, the energy density and the net-baryon density, by using a coarse graining procedure. Finally, we employ the Equation of State (EoS) to associate to these quantities the corresponding temperature and baryon chemical potential. For first studies in this respect see e.g. \cite{Bravina:1998pi,Bravina:1998it}.

The current limits of the chosen approach do not compromise the main goal of this study: highlighting the nature of the kinetic freeze-out as a continuous, dynamical process, by exploring the distribution of the kinetic freeze-out parameters at different collision energies. We focus on Au+Au reactions in the $0-5\%$ centrality class and extract temperature and baryon chemical potentials at midrapidity as a function of transverse momentum and as a function of rapidity. We focus on the most abundant hadron species and postpone a detailed analysis of the difference between hadron species to future follow-up studies. 

The structure of this article is as follows. In Section~\ref{sec:procedures} we explain the UrQMD model, the coarse graining approach and the extraction procedure in details. In Section~\ref{sec:results} we present our results on (T,$\mu_B$) values for different collision energies, fluctuations of the decoupling temperatures and chemical potentials and on the transverse momentum and rapidity dependence of kinetic freeze-out parameters. In Section~\ref{sec:conclusions}, we summarize the main findings of this study, we review its present limitations and we hint at possible further developments in future works.\\

\section{Description of the model}
\label{sec:procedures}
The present approach is based on the Ultra-relativistic Quantum Molecular Dynamics (UrQMD)~\cite{Bass:1998ca,Bleicher:1999xi} transport model. UrQMD is employed for two different purposes: to compute the time evolution of the average properties of the system, by exploiting a coarse-graining method, and to determine the space-time coordinates of the kinetic freeze-out coordinates of the hadrons. UrQMD itself is a hadron cascade model that simulates the dynamics of a heavy ion collision based on the covariant propagation of hadrons. Interactions are modelled via the excitation of color flux-tubes (strings) and by further elastic and inelastic interactions of the hadrons. For details, the reader is referred to ~\cite{Bass:1998ca,Bleicher:1999xi}.

The UrQMD coarse-graining method was developed in Refs.~\cite{Endres:2013daa,Endres:2014zua,Endres:2015fna,Endres:2015egk,Inghirami:2018vqd} and used successfully to explore and predict dilepton and photon production from GSI-SIS to RHIC energies as well as to provide underlying events for heavy quark studies. Here we employ the same approach and shortly summarize the main ingredients. In the coarse-graining method one reconstructs thermal parameters based on the approximation of the hadronic distribution function
$f(\textbf{\textit{x}},\textbf{\textit{p}},t)$ as
\begin{equation}
f(\textbf{\textit{x}},\textbf{\textit{p}},t)=\left\langle \sum_{h}
\delta^{(3)}\left(\textbf{\textit{x}}-\textbf{\textit{x}}_{h}(t)\right)
\delta^{(3)}\left(\textbf{\textit{p}}-\textbf{\textit{p}}_{h}(t)\right)\right\rangle,
\label{eq:distrib_function}
\end{equation} 
by performing averages over the total ensemble of hadrons produced in a large set of heavy ion collision events having the same $\snn$ energy.
These averages are done at each space point at fixed times (with respect to the UrQMD computational frame). The spatial grid for the coarse-graining procedure has a typical resolution of $0.8\,\fm$, except for collisions at $\sqrt{s_{NN}}=200\,\GeV$, for which we use a resolution of $1\,\fm$ to slightly reduce memory and disk space usage (see Table~\ref{table:overview}). More precisely, we evaluate the net-baryon four current $j^{\mu}_{\mathrm{B}}$ as
\begin{equation}
j^{\mu}_{\mathrm{B}}(\x,t)=\frac{1}{\mathrm{\Delta} V}\left\langle \sum\limits_{i=1}^{N_h \in \Delta V} B_i \frac{p^{\mu}_{i}}{p^{0}_{i}}\right\rangle,
\label{eq:j_avg}
\end{equation}
and the energy momentum tensor $T^{\mu\nu}$ as
\begin{equation}
T^{\mu\nu}(\x,t)=\frac{1}{\Delta V}\left\langle \sum\limits_{i=1}^{N_{h} \in \Delta V} \frac{p^{\mu}_{i} p^{\nu}_{i}}{p^{0}_{i}}\right\rangle,
\label{eq:Tmunu_avg}
\end{equation}
in which $\Delta V$ stands for the cell volume, $B_i$  and $p^{\mu}_i$ for the baryon number and the $\mu$ component of the four momentum of the hadron $i$, respectively, and the sums are done over all hadrons $N_{h}$. Adopting the Eckart's frame definition~\cite{Eckart:1940te}, we obtain the fluid four velocity $u^{\mu}$ from $j^{\mu}_{\mathrm{B}}$ as
\begin{equation}
u^{\mu}=\dfrac{j^{\mu}_{\mathrm{B}}}{\sqrt{j^{\xi}_{\mathrm{B}}{j_{\mathrm{B}}}_\xi}}=(\gamma,\gamma \vec{v}),
\label{eq:fluid_vel}
\end{equation}
with $u^{\mu}u_{\mu}=1$, $\gamma$ the Lorentz factor and $v$ the fluid velocity in natural units ($c=\hbar=1$). The baryon density $\rho_{\mathrm{B}}$ and the energy density $ \varepsilon$, as measured in the Local Rest Frame (LRF) of the fluid, can be obtained by a Lorentz transformation of the net-baryon current and of the energy momentum tensor as
\begin{equation} 
\rho_{\mathrm{B}} = j_{\mathrm{B,\,LRF}}^{0},\qquad \varepsilon = T^{00}_{\mathrm{LRF}}.
\end{equation} 

The temperature $T(\varepsilon,\rho_{\mathrm{B}})$ and the baryon chemical potential $\mu_{\mathrm{B}}(\varepsilon,\rho_{\mathrm{B}})$ are obtained by interpolation from a tabulated Hadron Resonance Gas EoS~\cite{Zschiesche:2002zr}, having consistently the same degrees of freedom as UrQMD. We accept a coarse-grained cell only, if it contains at least 100 particles (summing over all events), so to reduce the statistical fluctuations. Typically, we are able to determine the corresponding medium average bulk properties for more than $95\%$ of the kinetically frozen-out hadrons, except at $\sqrt{s_{NN}}=200\,\GeV$, where we drop at $\approx85\%$. In line with previous studies,  we rescale $\rho_{\mathrm{B}}$ and $\varepsilon$ before the interpolation step by a correction factor to compensate for the anisotropy of the system along the beam direction~\cite{Florkowski:2010cf,Endres:2014zua,Moreau:2019vhw}. However, this correction term is predominantly active only in the initial stages of the collision and not at the late times at which most of the kinetic freeze-out events happen. We verified for selected cases that its influence on the final results is negligible. 
The knowledge of the bulk properties of the system obtained by the coarse graining approach at different space-time points is then associated to the microscopic freeze-out distribution in space and time as given by UrQMD. 

\section{Results}
\label{sec:results}
We simulate central Au+Au collisions with impact parameter $b=0-3.4\,\fm$, roughly corresponding to the $0-5\%$ centrality class~\cite{Adamczewski-Musch:2017sdk,Abelev:2013vea}, from $\sqrt{s_{NN}}=2.4,\,\GeV$ to $\sqrt{s_{NN}}=200\,\GeV$, covering a range of energies relevant for the HADES~\cite{Agakishiev:2009am} at GSI, NA49~\cite{Blume:2004ci} at CERN and RHIC/BES~\cite{Odyniec:2013aaa} at BNL experiments. In Table~\ref{table:overview} we provide the details for the coarse graining simulations.  The lists of the kinetic freeze-out points have been obtained by running $10^4$ UrQMD events for each collision energy until $200\,\fm/c$. We consider the most abundant and significant hadron species, i.e. pions, kaons, protons, neutrons, lambdas and their antiparticles, including the feed-down of resonance decays.
\begin{table}[!ht]
	\footnotesize
	\begin{tabular}{| r | r | r | r | r | r | r |}
		
		\hline
		$\snn$ (GeV)&$N_{ev.}$&$\Delta t$ (fm/c)&$\Delta x$ (fm)&$N_{x,y}$&$N_z$&$t_{max}$ (fm/c)\\
		\hline
		$2.4$&$1.8\cdot10^6$&$0.5$&$0.8$&$70$&$200$&$80$\\
		$4.5$&$7.4\cdot10^5$&$0.5$&$0.8$&$80$&$250$&$90$\\
		$7.7$&$6.6\cdot10^5$&$0.5$&$0.8$&$80$&$250$&$90$\\
		$19.6$&$3.6\cdot10^5$&$0.5$&$0.8$&$86$&$276$&$100$\\
		$200$&$6.4\cdot10^4$&$0.5$&$1.0$&$200$&$402$&$200$\\		
		\hline	
	\end{tabular}
	\caption{List of the main parameters used in the UrQMD/coarse-graining numerical simulations. We report the values of the collision center-of-mass energy $\snn\,(\GeV)$, the number of events $N_{ev.}$, the time resolution $\Delta t\,(\fm/c)$, the spatial resolution $\Delta x\,(\fm)$, the number of cells along in the transverse plane ($N_{x,y}$) and in longitudinal direction ($N_z$), and the time $t_{max}\,(\fm)$  after the collision at which we stop the simulations.}
	\label{table:overview}
\end{table}

\subsection{Freeze-out time distributions, temperature and baryo-chemical potential variations on the decoupling hyper-surface}

\begin{figure}[ht!]
	\includegraphics[width=\linewidth]{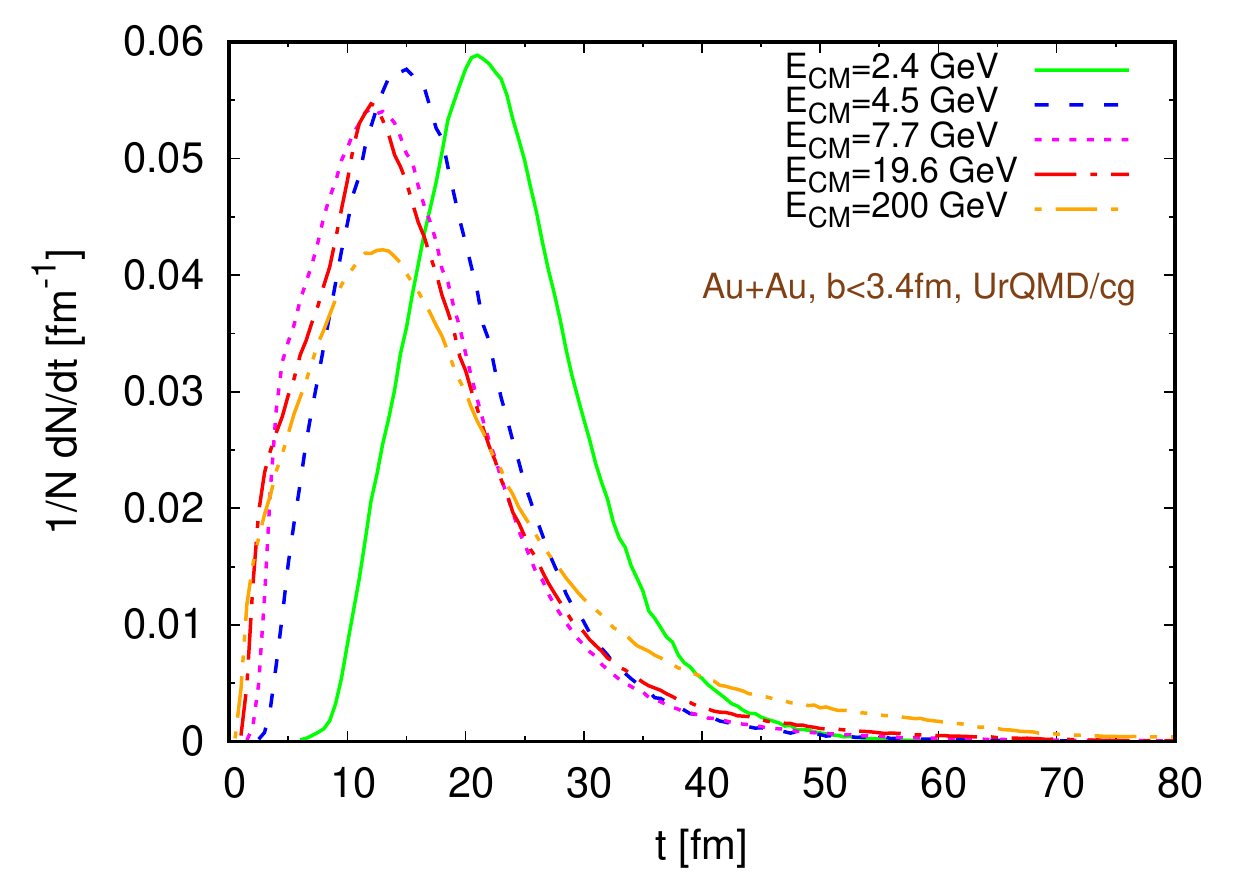}
	\caption{(Color online) Freeze-out time distribution of hadrons at midrapidity ($|y|<0.2$) for central Au+Au reaction at center-of-mass energies of $\snn=2.4, 4.5, 7.7, 19.6, 200$~GeV (full line, short dashed line, dashed line, long dashed-dotted line, dotted dashed line). The distributions are normalized to unity.} 
	\label{dNdtime_plot}
\end{figure}

To set the stage, we begin with the decoupling-time distribution defining the kinetic freeze-out. The studies focus on central rapidities ($|y|<0.2$) and the time $t$ is defined in the center-of-mass frame starting from the beginning of the collision. Fig.~\ref{dNdtime_plot} shows the decoupling probability (i.e. the normalized time distribution of the decoupling distribution) of the hadrons in central Au+Au reactions from 2.4~GeV to 200~GeV. The peak of the decoupling time is typically between 10 and 25 fm/c. The duration of the decoupling stage lasts typically $15-20$~fm/c (FWHM) (resulting in a damping rate $\Gamma(t_{\rm max})=30-40$~MeV) indicating that the kinetic freeze-out happens within a quite broad interval of time. It is interesting to note that the results are in line with the Kadanoff-Baym equation based analysis by Knoll \cite{Knoll:2008sc}. The position of the emission peak is governed to first approximation by two effects: I) the transition time of the initial nuclei and II) the expansion dynamics of the newly created matter. At low energies, the transition time provides the relevant scale, here the two initial nuclei will need at least the time span $d/\gamma_{\rm CM}\cdot v$ ($d$ is the diameter, $\gamma_{\rm CM}$ is the Lorentz gamma factor in the center-of-mass frame and $v$ is the velocity in the center-of-mass frame) to pass through each other, at $\snn=2.4$~GeV, this yields $d/\gamma_{\rm CM} \cdot v \approx 22$~fm/c as observed in the Figure. At higher energies, the initial nuclei are strongly Lorentz-contracted, here the dynamics of the meson dominated matter becomes the leading effect resulting in similar (and shorter) decoupling times with increasing energy. However, the tail of the distribution is more extended at the higher reaction energy, probably because of the larger particle multiplicity and a larger transverse Lorentz-boost\footnote{%
Since in our definition of kinetic freeze-out we include particle decays, the time dilation of the unstable particles lifetime observed in the computational frame can also extend significantly the tail of the distribution at high reaction energies. However, we expect that this effect is quite limited in the central rapidity region on which we are focusing.%
}.

\begin{figure}[ht!]
	\includegraphics[width=\linewidth]{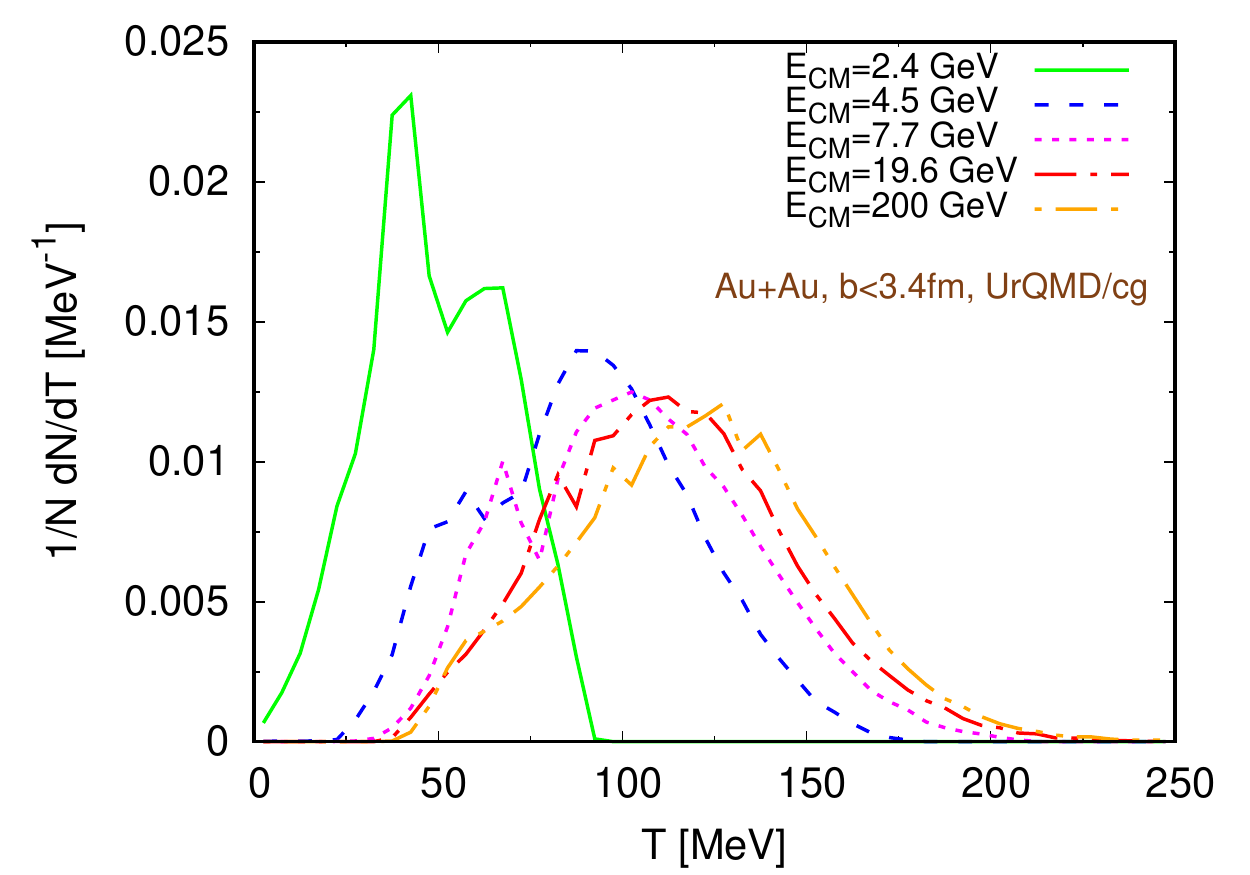}
	\caption{(Color online) Emission probabilities as a function of temperature at midrapidity ($|y|<0.2$) for central Au+Au reaction at center-of-mass energies of $\snn=2.4, 4.5, 7.7, 19.6, 200$~GeV (full line, short dashed line, dashed line, long dashed-dotted line, dotted dashed line). The distributions are normalized to unity.}
	\label{dN_dT}
\end{figure}

It is clear that decoupling probabilities in space and time can be transformed into a probability distribution for the temperature and baryo-chemical potential at the decoupling hyper-surface as calculated by the coarse graining procedure. We start with the analysis of the temperature distribution. To this aim, Fig.~\ref{dN_dT} depicts the normalized distributions of the temperatures at kinetic freeze-out in central Au+Au reactions from 2.4~GeV to 200~GeV (from left to right). The curves show clear maxima, however also a rather broad distribution. As expected from the chemical freeze-out curve, the peak kinetic emission temperature increases with increasing collision energy. Also for the kinetic freeze-out temperature, we observe that the peak temperature does not rise above a certain threshold of approx. 150~MeV, even at the highest energies. One also observes that some of the emission temperatures reach out to $T \approx 170-200$~MeV. This is due to the use of the hadronic equation of state in the present simulations as is adequate for a purely hadron based model. As an alternative, for such high temperatures we could have used an EoS based on a fitting to lattice QCD results\cite{He:2011zx,Borsanyi:2010cj}, merged, for $T$ below the critical temperature $T_c$, to the HG EoS in the limit of $\mu_{B}=0$. However, during a preliminary evaluation of this choice, we detected some artifacts when crossing $T_c$ at non-vanishing $\mu_{B}$. Since, most likely, the adoption of a purely hadron based model has a larger impact on the final results than the adoption of an EoS which does not affect the system evolution, but only the determination of the temperature of the kinetic freeze-out in the hottest cells, we prefer to use only the simple HG EoS, postponing the devising of a better EoS to future works.\\

\begin{figure}[ht!]
	\includegraphics[width=\linewidth]{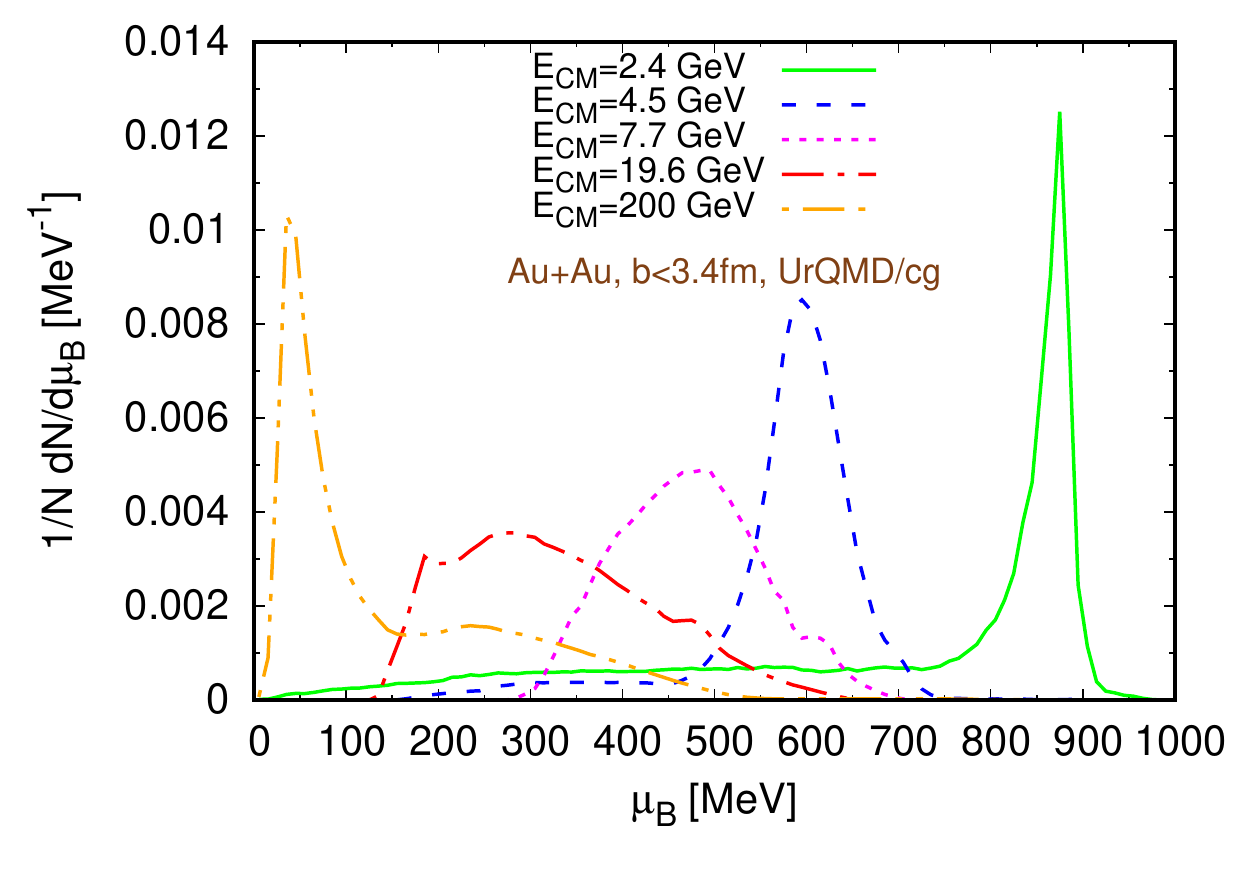}
	\caption{ (Color online) Emission probabilities as a function of baryo-chemical potential at midrapidity ($|y|<0.2$) for central Au+Au reaction at center-of-mass energies of $\snn=2.4, 4.5, 7.7, 19.6, 200$~GeV (full line, short dashed line, dashed line, long dashed-dotted line, dotted dashed line). The distributions are normalized to unity.}
	\label{dN_dmu}
\end{figure}

Next we turn to the emission probability as a function of the baryo-chemical potential in central Au+Au reactions from 2.4~GeV to 200~GeV (from right to left). Fig.~\ref{dN_dmu} shows the normalized distribution of the kinetic freeze-out points with respect to the baryon chemical potential $\mu_{\mathrm{B}}$. The curves show a pronounced peak structure, reflecting the baryon density in the late stage of the reaction. As expected, the distribution is peaked at high $\mu_{\mathrm{B}}$ for low collision energies and at low $\mu_{\mathrm{B}}$ for for high collision energies. Especially at intermediate energies, the distributions are rather broad due to the change from a baryon dominated system to a meson dominated system. 

Such a behaviour is expected from a kinetic freeze-out reflecting the complex dynamics of the system intertwined with the scattering cross sections of the hadrons, naturally leading to a kinetic freeze-out that is not only continuous in time, but also occurring in rather extended ranges of temperatures and densities. 

\subsection{Transverse momentum and rapidity dependence of the kinetic freeze-out parameters}

\begin{figure}[ht]
	\includegraphics[width=\linewidth]{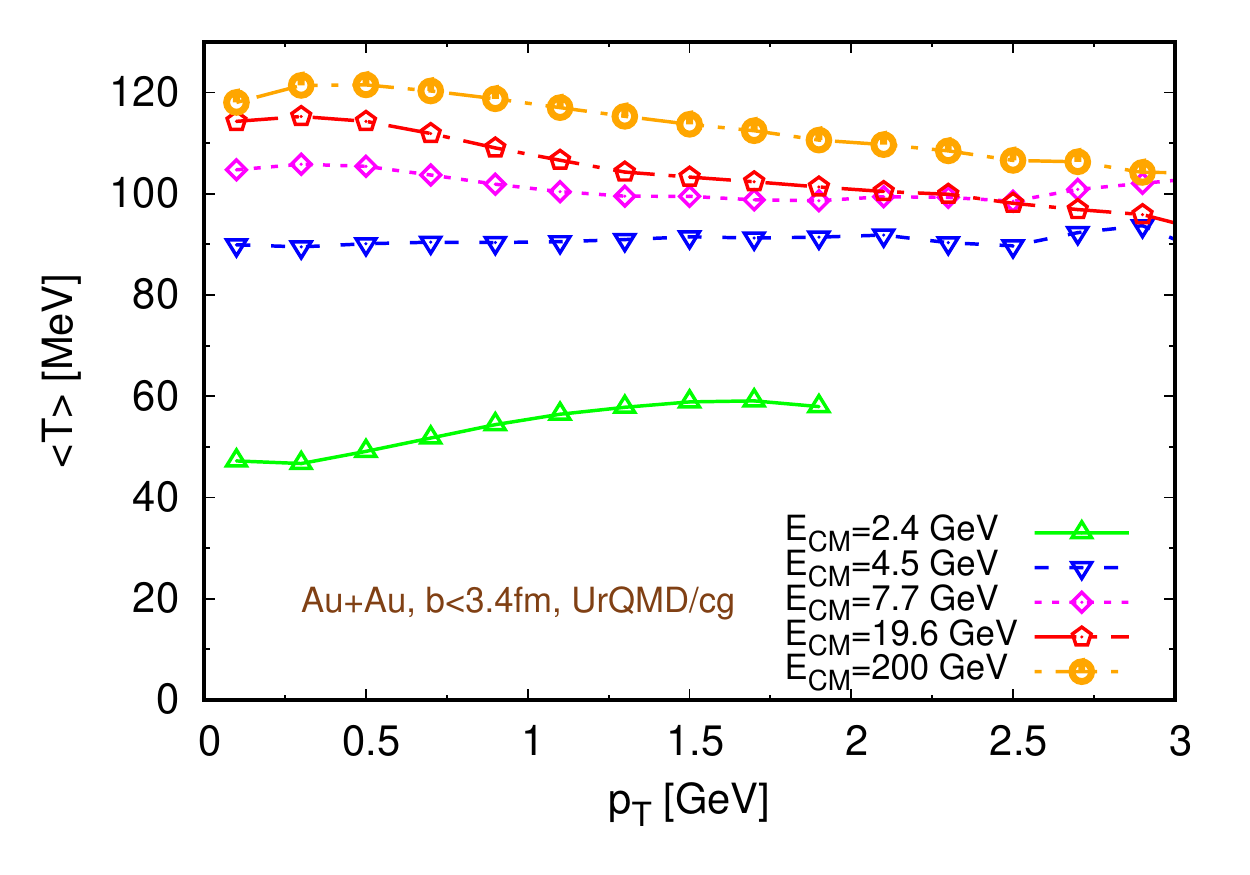}
	\caption{(Color online) Average kinetic freeze-out temperature $\langle T \rangle$ as a function of transverse momentum $p_T$ at midrapidity ($|y|<0.2$) for central Au+Au reaction at center-of-mass energies of $\snn=2.4, 4.5, 7.7, 19.6, 200$~GeV (full line, short dashed line, dashed line, long dashed-dotted line, dotted dashed line).}
	\label{T_pt}
\end{figure}

A question that arises now is whether and how the thermal freeze-out parameters are correlated with the rapidity or the transverse momenta of the hadrons. In Fig. ~\ref{T_pt} we explore the transverse momentum dependence of the average kinetic decoupling temperature at midrapidity ($|y|<0.2$) for central Au+Au reaction at center-of-mass energies of $\snn=2.4, 4.5, 7.7, 19.6, 200$~GeV. Generally, the dependence on the transverse momentum is rather weak. However, one can notice some interesting differences: At low collision energy, $\snn=2.4\,\GeV$, the average decoupling temperature tends to grow with increasing $p_T$, in contrast at higher energies, $\snn=19.6,\,200\,\GeV$, $\langle T \rangle$ decreases with increasing transverse momentum. This decrease at high collision energy indicates that the high $p_T$ hadrons emerge mainly from the outer cooled down regions~\cite{Reisdorf:1997fx,Schnedermann:1993ws} (they reach their high transverse momenta due to the substantial flow that has developed during the course of the evolution). At the lowest collision energy the situation is different, here a high transverse momentum hadron with a $p_T \approx 1-2$~GeV is typically produced only in the early (non-equilibrium) stages of the collision where initial nucleon-nucleon collisions with sufficient energy are available to reach such a high transverse momentum (as compared to the center-of-mass energy, being only $\snn=2.4\,\GeV$). 

\begin{figure}[ht]
	\includegraphics[width=\linewidth]{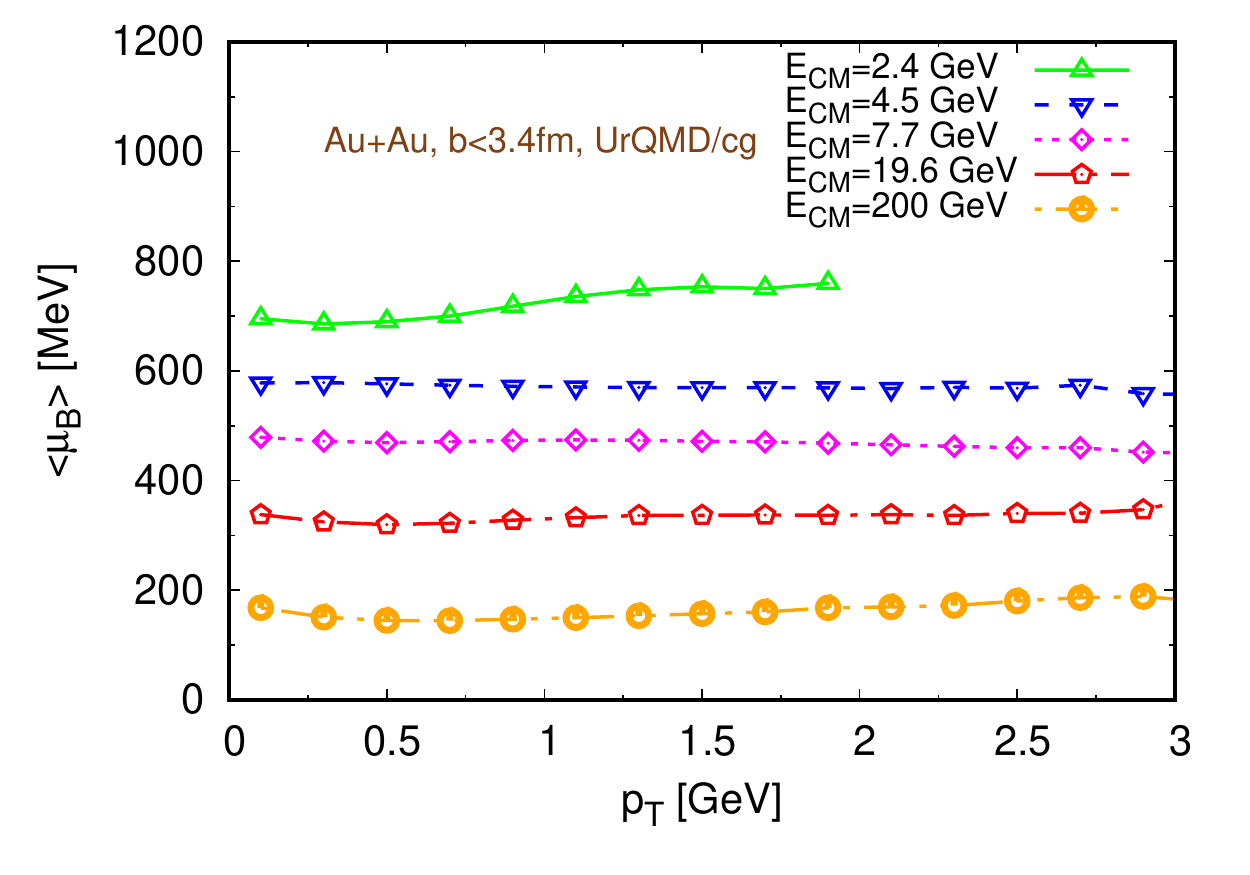}
	\caption{(Color online) Average baryo-chemical potential $\langle\mu_{\mathrm{B}}\rangle$ at kinetic freeze-out as a function of transverse momentum $p_T$ at midrapidity ($|y|<0.2$) for central Au+Au reaction at center-of-mass energies of $\snn=2.4, 4.5, 7.7, 19.6, 200$~GeV (full line, short dashed line, dashed line, long dashed-dotted line, dotted dashed line).}
	\label{mu_pt}
\end{figure}

Next we turn to Fig. ~\ref{mu_pt} and explore the transverse momentum distribution of the average baryo-chemical potential for the same reactions as above. Here we also observe only a weak dependence of the baryo-chemical potential as a function of transverse momentum. In line with our argument given above, we do observe a slight increase in the baryo-chemical potential with increasing $p_T$ for the lowest collision energy, which is consistent with the emission from an early reaction stage. 

\begin{figure}[ht]
	\includegraphics[width=\linewidth]{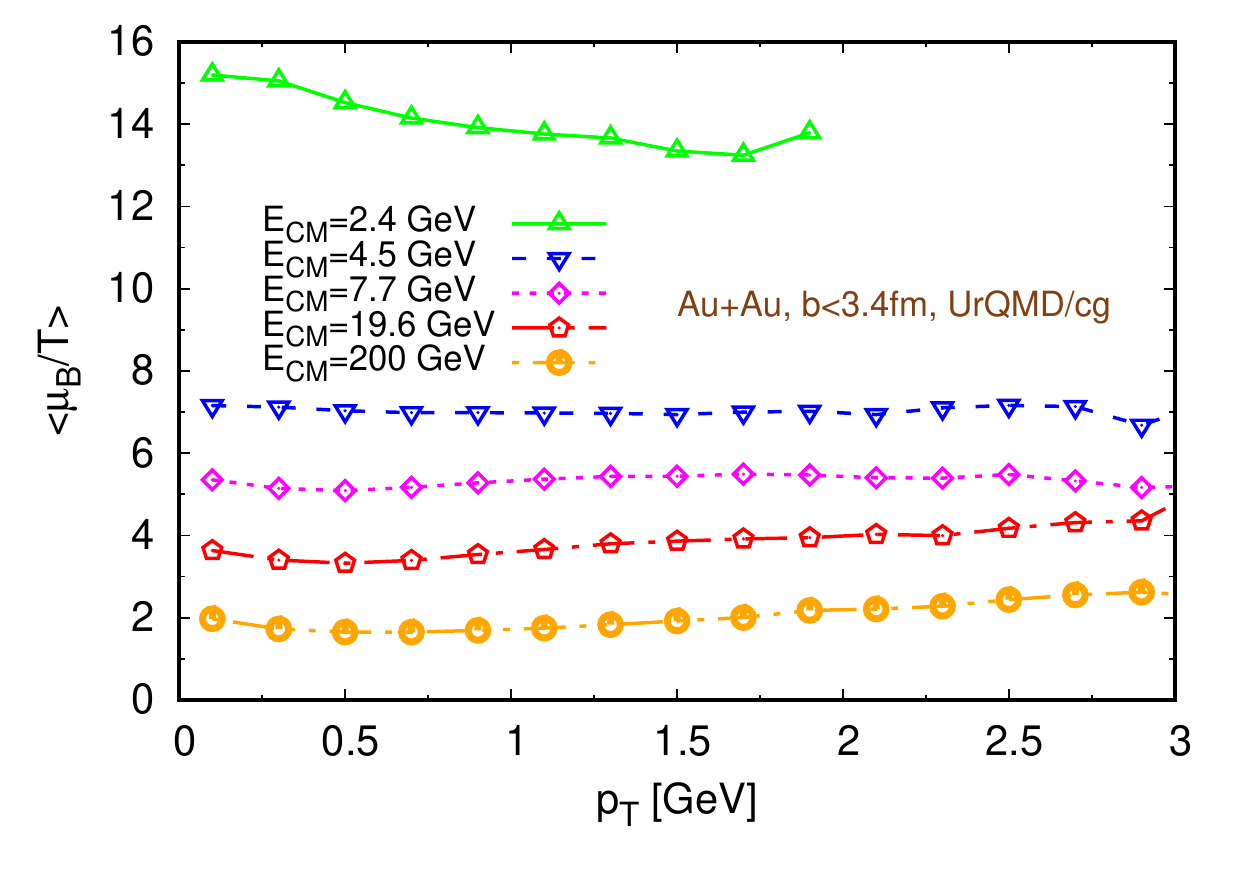}
	\caption{(Color online) Average of the ratio $\langle\mu_{\mathrm{B}}/T\rangle$ at kinetic freeze-out as a function of transverse momentum $p_T$ at midrapidity ($|y|<0.2$) for central Au+Au reaction at center-of-mass energies of $\snn=2.4, 4.5, 7.7, 19.6, 200$~GeV (full line, short dashed line, dashed line, long dashed-dotted line, dotted dashed line).}
	\label{muoverT_pt}
\end{figure}

We summarize these findings in Fig.~\ref{muoverT_pt} showing the $\langle\mu_{\mathrm{B}}/T\rangle(p_T)$ for the same reactions as above. Except for the lowest energy, we observe again a rather flat transverse momentum dependence. We conclude that while the freeze-out process is time dependent and continuous, we do not observe any sizable (bigger than 10\%) deviations of the average freeze-out temperature and baryo-chemical potential as a function of transverse momentum. 

\begin{figure}[ht]
	\includegraphics[width=\linewidth]{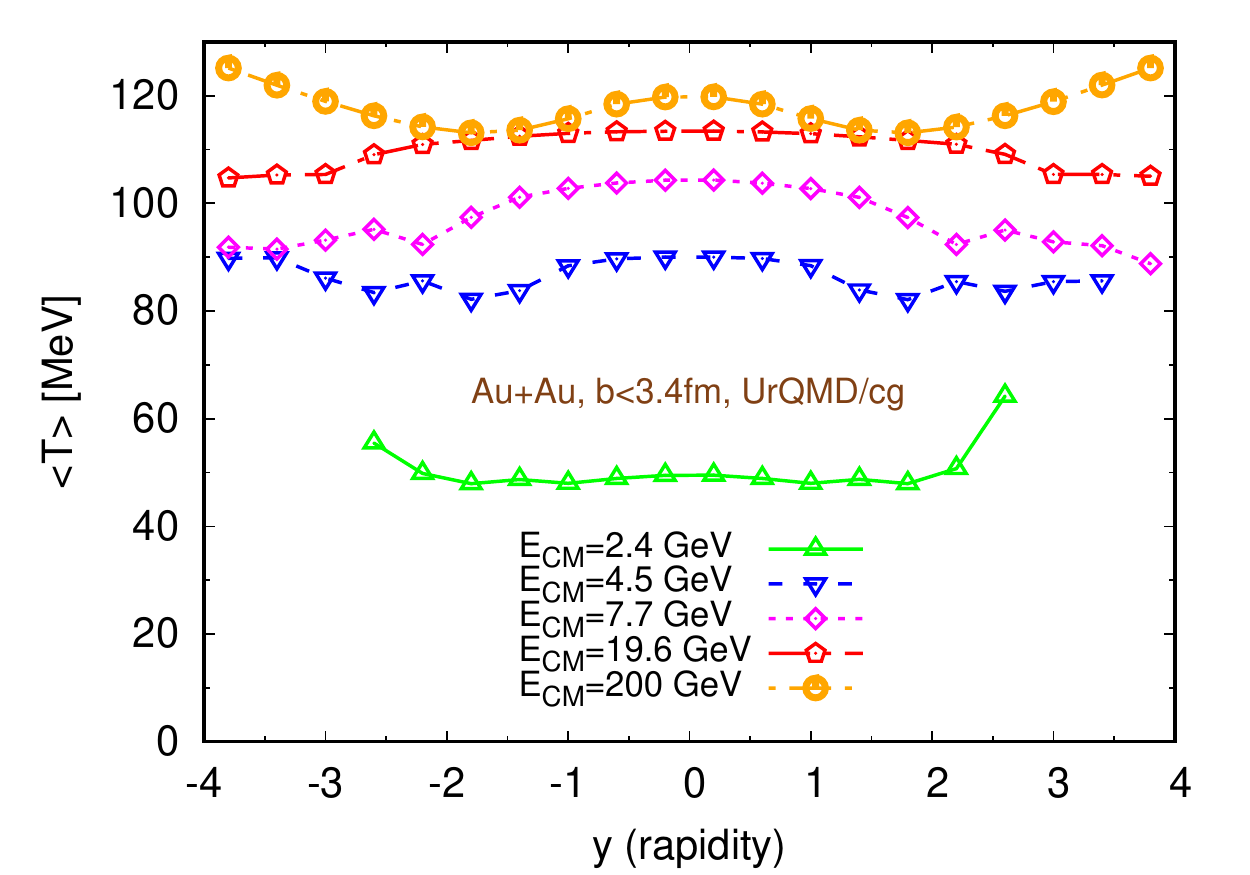}
	\caption{(Color online) Average kinetic freeze-out temperature $\langle T \rangle$ as a function of rapidity $y$ for central Au+Au reaction at center-of-mass energies of $\snn=2.4, 4.5, 7.7, 19.6, 200$~GeV (full line, short dashed line, dashed line, long dashed-dotted line, dotted dashed line).}
	\label{T_rap}
\end{figure}

\begin{figure}[ht]
	\includegraphics[width=\linewidth]{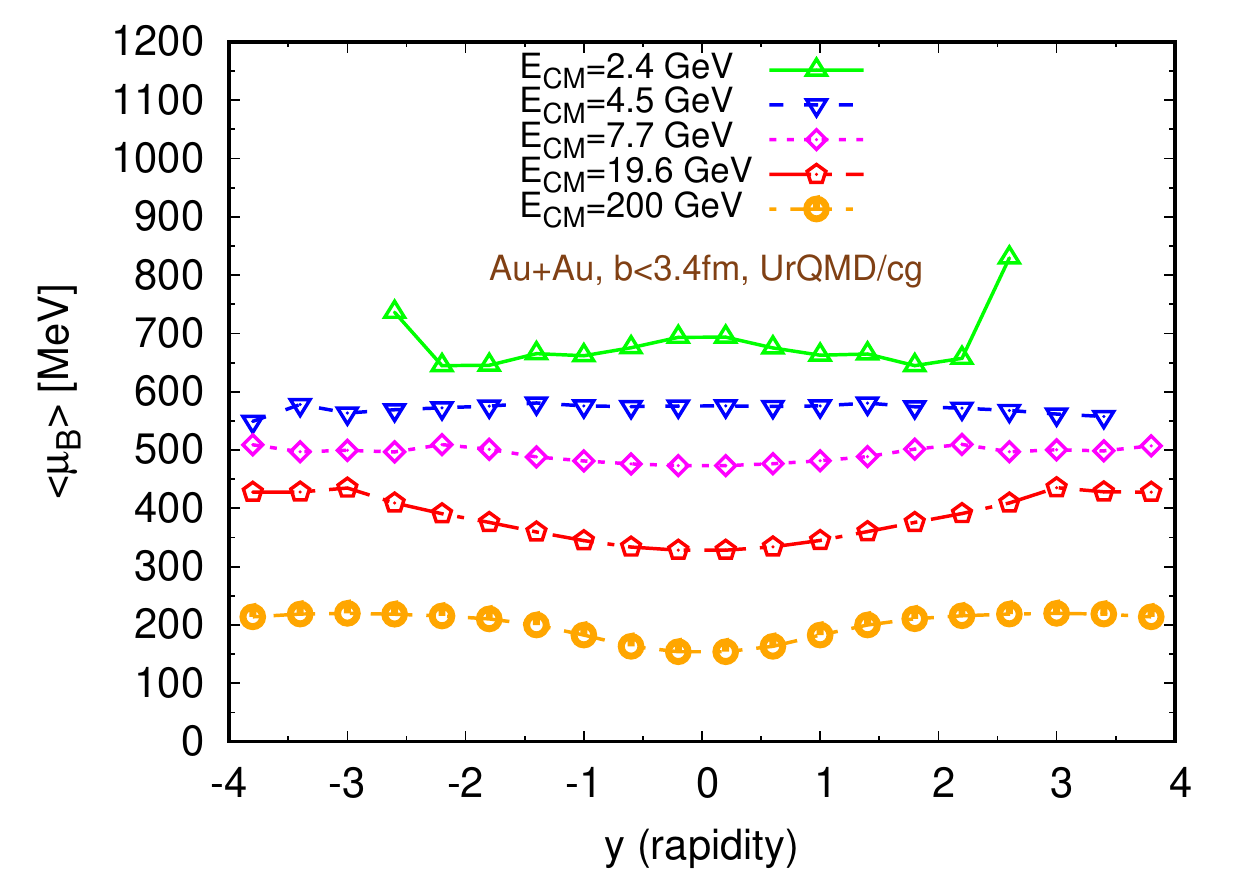}
	\caption{(Color online) Average baryo-chemical potential $\langle\mu_{\mathrm{B}}\rangle$ at kinetic freeze-out as a function of rapidity $y$ for central Au+Au reaction at center-of-mass energies of $\snn=2.4, 4.5, 7.7, 19.6, 200$~GeV (full line, short dashed line, dashed line, long dashed-dotted line, dotted dashed line).}
	\label{mu_rap}
\end{figure}

\begin{figure}[ht]
	\includegraphics[width=\linewidth]{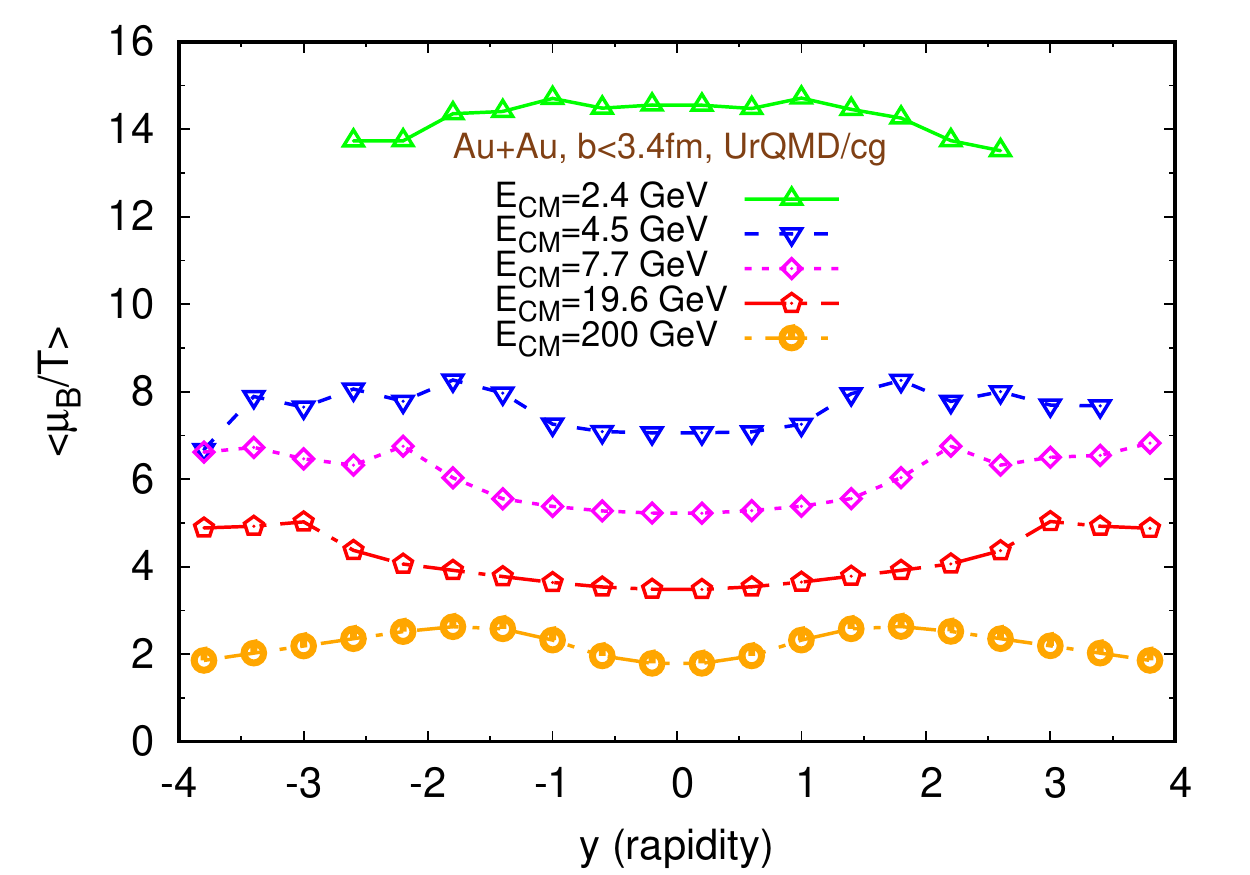}
	\caption{(Color online) Average of the ratio $\langle\mu_{\mathrm{B}}/T\rangle$ at kinetic freeze-out as a function of rapidity $y$ for central Au+Au reaction at center-of-mass energies of $\snn=2.4, 4.5, 7.7, 19.6, 200$~GeV (full line, short dashed line, dashed line, long dashed-dotted line, dotted dashed line).}
	\label{muoverT_rap}
\end{figure}

In the longitudinal direction one may expect stronger variations due to the change from the fireball region around midrapidity towards the fragmentation region in the forward and backward hemispheres. To explore this, Fig. ~\ref{T_rap} shows the dependence of the average temperature as a function of rapidity for central Au+Au reaction at center-of-mass energies of $\snn=2.4, 4.5, 7.7, 19.6, 200$~GeV. The kinetic freeze-out temperatures are again ordered by collision energy (increasing temperature with increasing beam energy) and show weak maxima in the central rapidity regions. For the baryo-chemical potential the curves are reversed and generally tend to show a minimum at central rapidities (except for the lowest collision energy) as shown in Fig.~\ref{mu_rap}. We again summarize our findings in Fig.\ref{muoverT_rap} and show $\langle\mu_{\mathrm{B}}/T\rangle(y)$. We notice that also this ratio tends to be quite stable, with a slightly smaller value in the central rapidity region and a mild enhancement at intermediate $|y|$. As in the case of the distributions with respect to the transverse momentum, the results present a clear hierarchy depending on the reaction energy, without substantial overlapping.

\subsection{Temperature and baryon chemical potential diagrams}
\begin{figure}[ht!]
	\includegraphics[width=\linewidth]{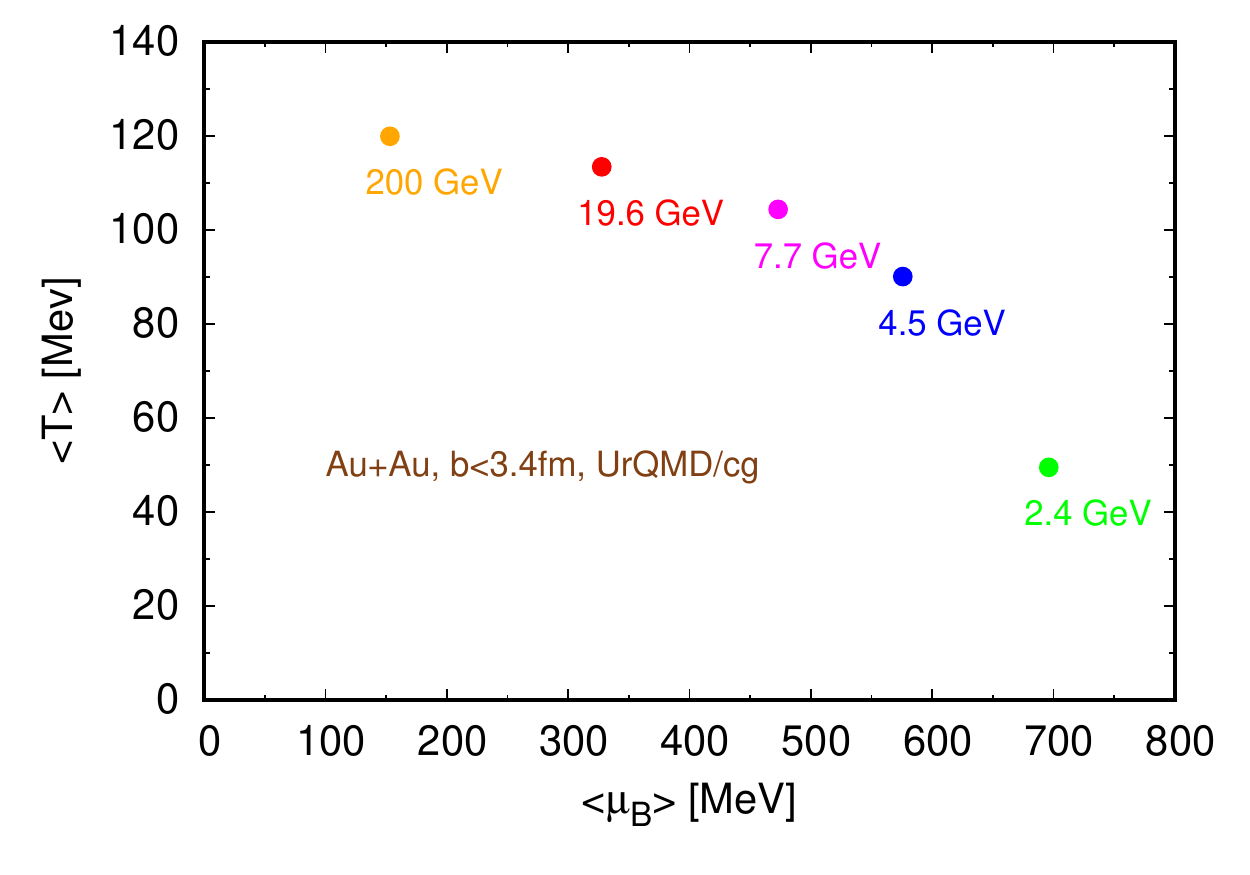}
	\caption{(Color online) Kinetic freeze-out temperature with respect to the baryon chemical potential in Au+Au reactions at different center-of-mass energies in the rapidity range $|y|<0.2$. } 
	\label{T_mu_plot}
\end{figure}

To allow for a simple comparison with the chemical freeze-out curve or Blast-Wave fits, we summarize the present results for the average kinetic decoupling temperature and chemical potentials in the $T-\mu_B$-diagram. In Fig.~\ref{T_mu_plot} we show the average temperature and baryon chemical potentials for the five reaction energies investigated in this study. The pairs of ($\langle T \rangle, \langle\mu_{\mathrm{B}}\rangle$) points follow a regular pattern: as the collision energy increases, $\langle T \rangle$ increases  and  $\langle\mu_{\mathrm{B}}\rangle$ decreases. The rate of variation of ($\langle\mu_{\mathrm{B}}\rangle,\,\langle T \rangle$) is very large at low reaction energies, while it becomes very mild between $\snn=19.6\,\GeV$ and $\snn=200\,\GeV$. 

\begin{figure}[ht!]
	\includegraphics[width=\linewidth]{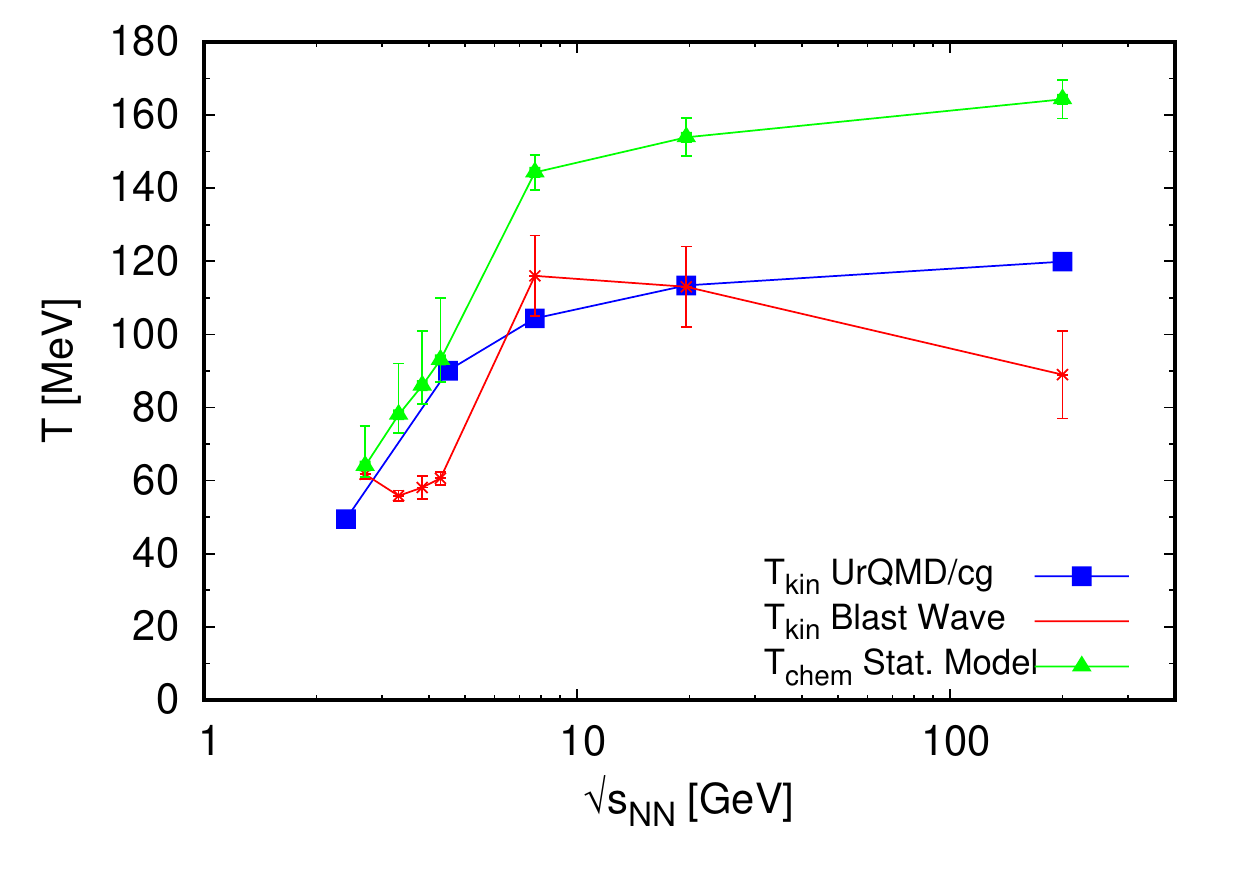}
	\caption{(Color online) Comparison between the average kinetic freeze-out temperature determined in the present study, the kinetic freeze-out temperature obtained from Blast-Wave model fits (Refs.~\cite{Adamczyk:2017iwn,Abelev:2008ab,Rode:2018hlj}) and the chemical freeze-out temperatures obtained from the Statistical Hadronization model fits (Ref.~\cite{Adamczyk:2017iwn}, table VIII, GCER, and Ref.~\cite{,Andronic:2005yp}), with respect to the reaction energy. The calculation and the data refer to central Au+Au reactions. We converted the reaction energies of fixed target experiments from $E_{lab}$ to $\sqrt{s_{NN}}$.}
	\label{T_vs_collen}
\end{figure}

In Fig.~\ref{T_vs_collen} we compare our results with those of previous studies. In particular, we compare the average kinetic freeze-out temperature determined in the present work with the kinetic freeze-out temperature obtained from Blast-Wave model fits and with the the chemical freeze-out temperature according to the Statistical Hadronization model. The data regarding the kinetic freeze-out temperature at $\snn=2.7,\, 3.32, \, 3.84$ and $4.3\,\GeV$ are taken from Ref.~\cite{Rode:2018hlj} and are based on the analysis of pions and protons with a double fit, first with respect to transverse momentum spectra at midrapidity ($|y|<0.05$), then with respect to the pseudorapidity, using a non-boost invariant Blast-Wave model, up to a maximum value which depends on the beam rapidity, which, of course, in turn depends on the collision energy. We refer to  Ref.~\cite{Rode:2018hlj} for the details. The data of the kinetic freeze-out at $\snn=7.7$ and $19.6\,\GeV$ come from Ref.~\cite{Adamczyk:2017iwn} and they have been obtained by a simultaneous fit with a Blast-Wave model with $|y|<0.1$ of $\pi^\pm$ ($0.5<p_T<1.3\,\GeV$), $K^\pm$ ($0.24<p_T<1.4\,\GeV$), $p$ and $\bar{p}$ ($0.4<p_T<1.3\,\GeV$). The authors of this study excluded other particle species to avoid the consequent implicit assumption that all hadrons share the same kinetic freeze-out temperature. The authors also imposed limits on the transverse momentum selected for the fits.  On the low $p_T$ end, this restriction was motivated by the issues with the resonance decays, while, on the high $p_T$ end, the hydrodynamic models underlying the Blast-Wave model is not adequate to describe hard processes~\cite{Adamczyk:2017iwn}. The data at $\snn=200\,\GeV$ are taken from Ref.~\cite{Abelev:2008ab} and also refer to a fit with a Blast-Wave model of $\pi^\pm$, $K^\pm$, $p$ and $\bar{p}$ at midrapidity ($|y|<0.1$), without considering the pion spectra for $p_T<0.5\,\GeV$. Our model of kinetic freeze-out incorporates the resonance feed-down, therefore their contribution should be quantitatively assessed for a detailed comparison between the two models, which, nevertheless, is out of the scope of the present work. Given the small abundance of hadrons at high $p_T$ with respect to those at low $p_T$, we are less concerned by a possible bias introduced by them. The data regarding the chemical freeze-out temperature between $\snn=2.7$ and $4.3\,\GeV$ are taken from Ref.~\cite{Andronic:2005yp}, while at the remaining reaction energies they are from Ref.~\cite{Adamczyk:2017iwn}. Let us first compare our kinetic freeze-out temperatures with the chemical freeze-out temperatures from the Statistical Model fits. We observe that at low collision energies ($\snn=7\,\GeV$), kinetic and chemical freeze-out are only separated by a small temperature difference on the order of 5-10~MeV, nicely consistent with a very short duration of the expansion phase. At energies above $\snn=7\,\GeV$ the chemical freeze-out temperature is substantially above the kinetic decoupling temperature ($\Delta T>40-50$~MeV). This indicates a rather strong expansion flow of the system from chemical to kinetic freeze-out at high energies. If we compare the kinetic freeze-out temperature from the present study to the kinetic freeze-out temperatures obtained from Blast-Wave fits, we observe that, apart from the point at $\snn=19.6,\GeV$, there is a tension between the results of the two approaches, with the Blast-Wave fits suggesting a substantially lower kinetic freeze-out temperature than obtained in the present study, in particular at high collision energy. We relate this difference to the hadronic dynamics that leads to weaker transverse expansion than observed in the data. For the present investigation we employ UrQMD without a hydrodynamic/QGP stage to avoid to introduce an additional parameter, the ``particlization'' temperature~\cite{Huovinen:2012is}, whose proper exploration would require a rather strong computational effort, made heavier by the longer time needed to run UrMQD in hydbrid mode compared to cascade mode. At low collision energy, the discrepancy might be due to the exclusion of the hadrons with low $p_T$ in the Blast-Wave fits. The inadequacy of our chosen EoS to describe a system out of chemical equilibrium might introduce a bias, as well. Further investigations to understand the differences between our results and those coming from the Blast-Wave model will be addressed in a future study, probably including a careful evaluation of the bias introduced in the fits by the selection of the $p_T$ intervals and the adoption of a different EoS.

\begin{figure}[ht!]
	\includegraphics[width=\linewidth]{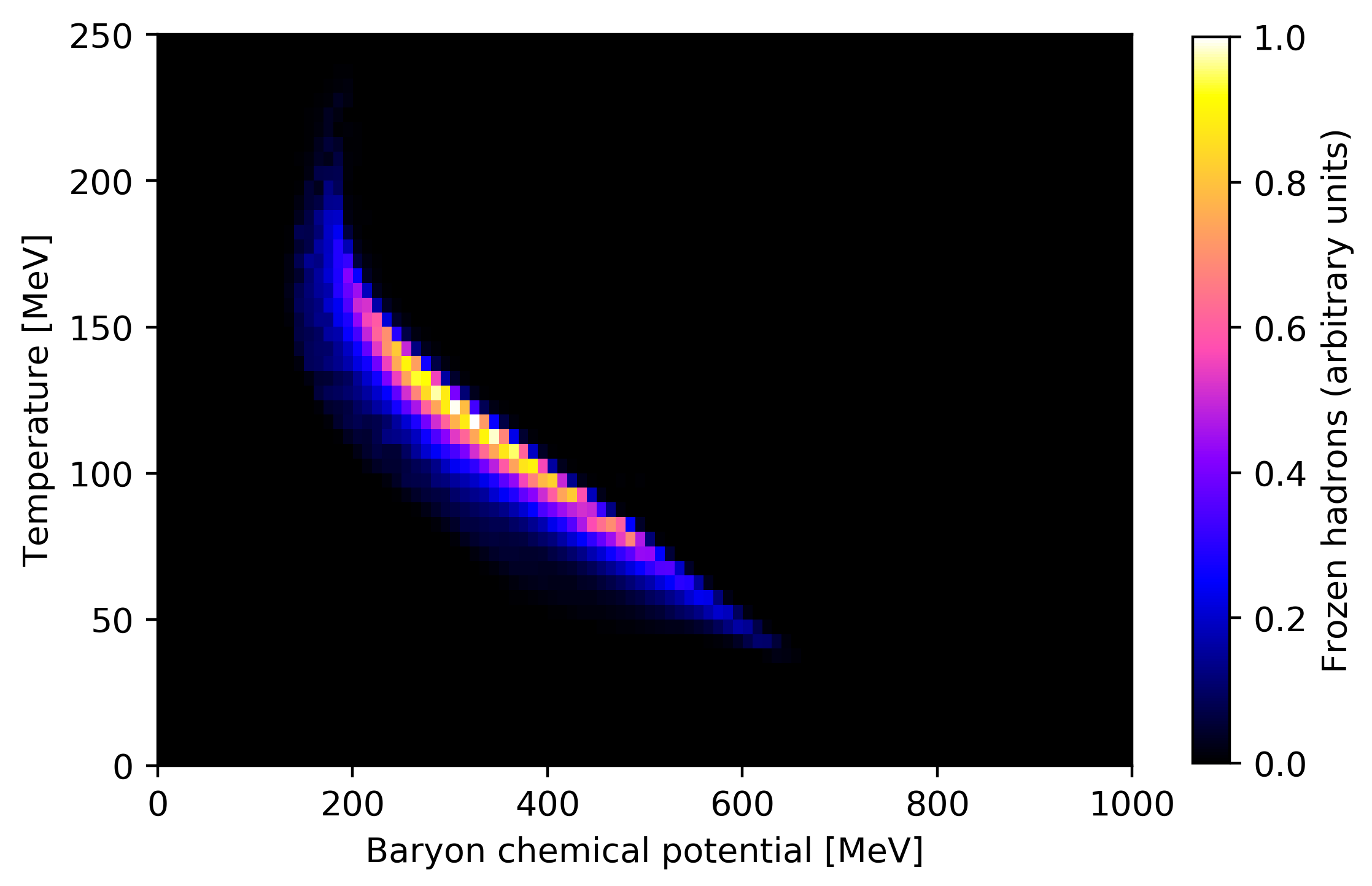}
	\caption{(Color online) Profile of the kinetic freeze-out temperature and baryon chemical potential at $|y|<0.2$ for Au+Au collisions at  $\snn=19.6\,\GeV$.}
	\label{phase_diagram_19_6}
\end{figure}

Under this perspective, one should not forget that the representation of the kinetic freeze-out as a single point in the phase diagram is indeed a convenient way to summarize its key properties, but, at the same time, it is also an oversimplification. For example, Fig.~\ref{phase_diagram_19_6} shows the density of the kinetic freeze-out parameters in the ($T,\mu_{\mathrm{B}}$) plane for central Au+Au reactions at $\snn=19.6\,\GeV$. One clearly observes that different parts of the system decouple at different ($T,\mu_{\mathrm{B}}$) points. In addition a correlation between $\langle\mu_{\mathrm{B}}\rangle$ and $\langle T \rangle$ is present. Such a spread in parameter space is at the moment not included in the present Blast-Wave fits and might yield different results than in the standard Blast-Wave approach.

\section{Summary and conclusions} 
\label{sec:conclusions}
In this work we studied the kinetic freeze-out process with the UrQMD/coarse-graining approach~\cite{Bass:1998ca, Bleicher:1999xi,Endres:2014zua}. First, we performed a large series of UrQMD simulations to compute the average temperature and baryon chemical potential of the system during its evolution on a coarse-grained grid. Then we determined the time and the position of the points of the last interaction of the most abundant hadron species. These space-time points of last interactions, which include both scatterings and decays by strong interaction, are what we defined as kinetic freeze-out hyper-surface. Afterwards, we associated to these last interaction points the corresponding values of the coarse-grained cell in which they were located. We focused on Au+Au collisions in the centrality class $0-5\%$, i.e. with a Glauber model impact parameter $b<3.4\,\fm$, considering five reaction energies: $\snn=2.4,\,4.5,\,7.7,\,19.6$ and $200\,\GeV$. 

We evaluated the probability distributions for particle emission in time and the probability for the emission at a given temperature and chemical potential. In general, we found that these distributions are rather broad. These results are consistent with the concept of kinetic freeze-out as continuous process which happens at very different space-time points, due to the complex dynamics of the system and due to the different (and energy dependent) cross sections of the hadrons. With increasing collision energy, the average freeze-out times tend to decrease, the average freeze-out temperatures become higher and the average baryon chemical potential decreases. We evaluated also how the average baryon chemical potential, the average temperature and the average of the ratio between the two vary with respect to the transverse momentum and to the rapidity. We found that these average values are essentially independent of rapidity and transverse momentum. 

Finally, we presented the set of the average temperature and baryon chemical potential points at kinetic freeze-out at the various collision energies under investigation, comparing our results with those coming from Blast-Wave model fits for the kinetic freeze-out temperature and from the Statistical Hadronization model for the chemical freeze-out temperature. We found that the kinetic freeze-out points in the ($T, \mu_{\mathrm{B}}$) plane follow a regular pattern, from higher to lower baryon chemical potential and from lower to higher temperature as the reaction energy grows, similar to the curve described by the chemical freeze-out points, albeit with different values. We observe some deviations between the results from the Blast-Wave fits, in which the kinetic freeze-out temperature at high collision energy seems to slightly decrease, which might be interesting to investigate more into depth in future studies. Moreover, we found also a disagreement in the low collision energy region which might be important to understand, giving the rapid variation of the key thermodynamical properties of the system in that reaction energy region, slightly below the bottom end of the BES-II program~\cite{Yang:2017llt} and in the range of the  upcoming FAIR~\cite{Friman:2011zz} and NICA~\cite{Syresin:2019vzo} facilities. We concluded by showing a density plot of the freeze-out parameters at $\snn=19.6\,\GeV$ to provide the evidence that the common representation of the kinetic freeze-out as a single, well defined point in the phase diagram hides its real nature as a continuous process across many different thermodynamical conditions.

\section*{Acknowledgments} 
We gratefully acknowledge Stephan Endres for providing the coarse-graining
numerical code which served as a basis for the present work. 
We sincerely thank the Referees for helping us in improving the quality of the manuscript. G. Inghirami is supported by the Academy of Finland, Project no. 297058. P. Hillmann acknowledges support by the GSI in cooperation with the John von Neumann Institute for Computing; she also acknowledges support from the HGS-HIRe and FIGSS graduate schools. B. Tom\'a\v{s}ik acknowledges support by the grant No. 17-04505S from the Czech Science Foundation. The computational resources were provided by the Center for Scientific Computing (CSC) of the Goethe University Frankfurt and by the Frankfurt Institute for Advanced Studies (FIAS). This work was supported by the COST Action CA15213 (THOR).
\bibliography{bibliography}
\end{document}